\documentclass[preprint,proceedings]{rmaa}

\def\hMsun{$h^{-1}{\ }{\rm M_{\odot}}$}
\def\ea{et~al.~}                            

\newcommand{\D}{\discretionary{}{}{}}

\SetYear{2002}
\SetConfTitle{Galaxy Evolution: Theory and Observations}

\title{Filaments in Warm Dark Matter}

\author{
  Alexander Knebe\altaffilmark{1}}

\altaffiltext{1}{Centre for Astrophysics \& Supercomputing,
                 Swinburne University, Mail\# 31, PO Box 218,
                 Hawthorn, VIC 3122, Australia
                 (aknebe\D{}@astro.\D{}swin.\D{}edu.\D{}au).}

\suppressfulladdresses

\listofauthors{A. Knebe}
\indexauthor{Knebe, A.}

\addkeyword{H~II regions}
\addkeyword{ISM: Jets and outflows}
\addkeyword{Stars: Pre-main sequence}
\addkeyword{Stars: Mass loss}

\begin{document}
\maketitle 

\boldabstract{Using Warm Dark Matter rather then Cold Dark Matter we
              are able to solve the over-production of satellite
              galaxies orbiting in galactic haloes. Moreover, we can
              show that this model leads to interesting new results in
              terms of the history and large-scale distribution of
              low-mass objects as well as the formation of filamentary
              structures when being compared to the standard CDM
              scenario.}

CDM models have been very successful in reproducing the large scale
structure properties of the universe. However, they have lately been
facing a state of crisis because of apparent discrepancies between
high resolution $N$-body simulations and observations on galaxy
scales. One can divide these problems into two categories: the
cuspiness of typical L$_\star$ galaxy halos on the one hand, and the
dearth of dark matter satellites in these very same halos on the other.

One suggested solution to these problems is the introduction of Warm
Dark Matter (e.g. Bode, Ostriker \& Turok 2001, and references
therein). We ran a series of such $N$-body simulations focusing the
comparison to the standard CDM model on the formation and merger
histories of the two most massive dark matter halos found in those
runs (Knebe~\ea 2002). The results show that WDM is able to reduce the
number of satellites within such halos. In addition to that we find
that halos in the mass range $M \in [10^{10},10^{11}]$\hMsun\
preferentially form along the filamentary structures as well as at
lower redshifts. Our interpretation is that in the WDM model the
filaments fragment as opposed to the standard hierarchical structure
formation scenario in CDM.

\begin{figure}[!t]
  \includegraphics[width=\columnwidth]{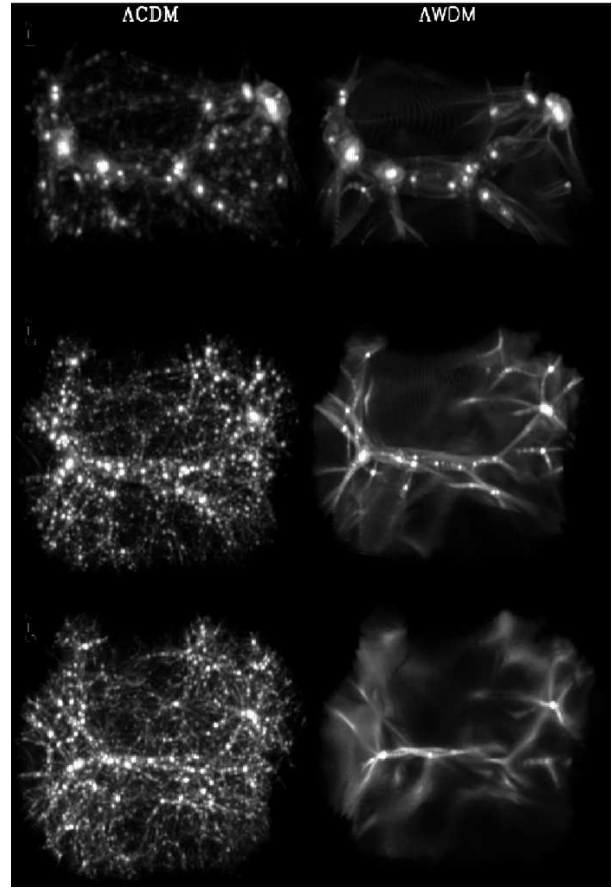}
  \caption{Evolution of a single filament in CDM and WDM.
           The redshifts are $z=1$ (top), $z=3$ (middle),
           and $z=5$ (bottom).}
  \label{Figure}
\end{figure}

To clarify this finding we re-simulated one of the filaments with much
higher mass resolution ($m_p=10^7$\hMsun). The time evolution of that
filament is presented in Fig.~\ref{Figure}. Despite the obvious
differences at redshift $z$=5 the filaments look astonishingly similar
at $z$=1. And for WDM we clearly observe individual (low-mass) halos
appearing during the course of the filament evolution. To further
investigate and quantify the formation history of objects within the
filament we performed a standard friends-of-friends (FOF)
analysis. Each individual FOF halo was then either tagged 'old' or
'new' depending on its presence at the previous redshift. However, a
halo was only marked 'old' when its progenitor was at least 1/3 third
of the current halo mass.  We are able to show that in WDM a large
fraction of objects is formed within the filament whereas in CDM
nearly all halos are already in place at redshift $z=5$. This can be
easily understood if the WDM filament itself fragments and hence forms
halos in a top-down fashion.

\begin{thebibliography}

\bibitem{Knebe02} 
  Knebe A., Devriendt J., Mahmood A., \& Silk J.,
  2002, MNRAS, 329, 813

\bibitem{Bode01} 
  Bode P., Ostriker J.P., Turok N., 
  2001, ApJ, 556, 93


\end{thebibliography}
\end{document}